\documentclass[twocolumn,prl]{revtex4}
\usepackage{graphicx,epsfig,amssymb}
\usepackage{amsmath}
\usepackage{graphicx}
\usepackage{amsfonts}
\usepackage{amssymb}

\begin{document}

\title{Projective Quantum Monte Carlo Method for the
Anderson Impurity Model\\ and its Application to Dynamical
Mean Field Theory}
\author{M. Feldbacher,$^{1}$ K.\ Held,$^{1}$  and F.\ F.\ Assaad$^{2}$}
\affiliation
{$^1$Max-Planck-Institut f\"ur Festk\"orperforschung, Heisenbergstra\ss
e 1, D-70569 Stuttgart\\ $^2$Universit\"at W\"urzburg, Institut f\"ur Theoretische Physik I, Am
Hubland, 97074 W\"urzburg}
\begin{abstract}
We develop a projective quantum Monte Carlo algorithm 
of the Hirsch-Fye type for  obtaining ground state properties
of the Anderson impurity model. 
This  method is employed
to solve the  self-consistency equations of dynamical mean field theory.
It is shown that the approach converges rapidly to the
ground state so that reliable zero-temperature results are obtained.
As a first application, we study the  Mott-Hubbard metal-insulator
transition of the one-band Hubbard model, reconfirming the 
numerical renormalization group results.
\end{abstract}


\date{\today}
\maketitle

In recent years, we have seen a revival of interest in Kondo-like physics, in
particular in quantum dot systems \cite{Kouwenhoven01}, adatoms on surfaces \cite{adatoms}, and in connection with
the dynamical mean field theory (DMFT) \cite{DMFT1,DMFT2}. The underlying
microscopic model, the Anderson impurity model (AIM), can be solved exactly by
Bethe ansatz \cite{Betheansatz} for the special case of a linear dispersion
relation. But in general, such a solution is only possible numerically. Two
sophisticated numerical techniques have been established in the literature: the
numerical renormalization group (NRG) \cite{NRG} and the Hirsch-Fye quantum
Monte Carlo (HF-QMC) \cite{HFQMC} method.

In his pioneering work \cite{NRG}, Wilson applied the renormalization group
concept to the Kondo problem: High energy degrees of freedom are
systematically integrated out and one obtains an excellent description of the
low energy degrees of freedom, which was not possible before since
perturbation theory for the AIM breaks down at low temperatures and energies.
Confirming earlier scaling ideas of Anderson \cite{Anderson}, these NRG
calculations show the formation of a low-energy Kondo singlet between the AIM
localized and conduction electrons and an associate Abrikosov-Suhl
resonance in the spectrum.

The same physics can be described by the HF-QMC algorithm. This method 
solves the AIM numerically by discretizing the imaginary time
interval $[0,\beta]$
($\beta=1/T$: inverse temperature) into steps of size $\Delta\tau$ and by
mapping the interacting AIM onto a sum of non-interacting
models via the Hubbard-Stratonovich transformation.

The two approaches require the extrapolation of a discretization parameter,
the logarithmic energy mesh parameter $\Lambda\rightarrow1$ (NRG) and
$\Delta\tau\rightarrow0$ (HF-QMC), and have some limitations: The NRG effort
grows exponentially with the number of localized AIM orbitals $M$, restricting
the algorithm to $M \leq2$, which prevents the usage of NRG for more realistic
calculations where one would like to include, e.g., $M=5$ or $7$ orbitals for
3$d$ and 4$f$ systems, respectively. The HF-QMC on the other hand scales like
$M^{2} (\beta/\Delta\tau)^{3}$ and, hence, quickly becomes too expensive in
CPU time if temperature is decreased. This is a severe restriction since
interesting many-body physics often occurs at low temperatures.

In this Letter, we introduce a new projective quantum Monte Carlo (PQMC) method
for studying the zero-temperature AIM in the thermodynamic limit. 
Similar to projective approaches in lattice models \cite{PQMC},
it relies on the idea that ground state properties are more
efficiently obtained by filtering out  -- via projection along the imaginary
time axis -- the ground state wave function 
from a \textit{suitably} 
chosen trial wave function.
For the AIM, however, an efficient algorithm
needs to be of the Hirsch-Fye type, in contrast
to the lattice case \cite{PQMC}.
In a second step, we incorporate this  algorithm as an impurity
solver within DMFT. In passing, we provide for a mandatory
confirmation of the T=0 DMFT scenario for
the  Mott-Hubbard  transition.


{\it PQMC algorithm --}
Our starting point is the AIM Hamiltonian
\begin{eqnarray}
  \!H_{\text{AIM}}&\!=\!&H_{0}+Un_{f\uparrow}n_{f\downarrow},\\
    H_{0}&\!=\!&\!\sum_{\sigma}\!\epsilon_{f}n_{f\sigma}\!+\!\sum_{k\sigma}\!V^{\phantom{\dagger}}_{k {\sigma}}(c_{k{\sigma}}^{\dagger}f^{\phantom{\dagger}}_{\sigma}\!+\!{\rm h.c.})+ \!\sum_{k  \sigma}\!\varepsilon^{\phantom{\dagger}}_{k}c_{k\sigma}^{\dagger}c^{\phantom{\dagger}}_{k\sigma}.\nonumber
\end{eqnarray}
Here, $f^{\phantom{\dagger}}_{\sigma}$($f^{\dagger}_{\sigma}$) create (annihilate) impurity electrons with spin $\sigma$ which interact
via $U$ and have a level energy $\epsilon_f$; 
$n_{f\sigma}=f^{\dagger}_{\sigma}f^{\phantom{\dagger}}_{\sigma}$.
These orbitals hybridize via $V^{\phantom{\dagger}}_{k {\sigma}}$
with a conduction band ($c_{k}^{\dagger}$),  having a
dispersion $\varepsilon_{k}^{\phantom{\dagger}}$.

Let us consider a trial wave function 
$\left|  \Psi_{T}\right\rangle$ which is
non-orthogonal to
the (non-degenerate) ground state  $|\Psi_{0}\rangle$ of the AIM.
By means of the $\theta$-
projector $\sim \exp(-\theta H_{\text{AIM}}/2)$,
we can filter out  $|\Psi_{0}\rangle$ from $\left|  \Psi_{T}\right\rangle $ and
calculate ground state expectation values of an 
arbitrary operator ${\cal O}$:
\begin{eqnarray}
\!\langle\mathcal{O}
\rangle_{0}&\!=\!&\frac{\langle\Psi_{0}|\mathcal{O}|\Psi_{0}\rangle}{\langle\Psi_{0}
|\Psi_{0}\rangle}\\
&=&\lim_{\theta\rightarrow\infty}\frac
{\left\langle \Psi_{T}\right|  e^{-\theta H_{\text{AIM}}/2}\mathcal{O}%
e^{-\theta H_{\text{AIM}}/2}\left|  \Psi_{T}\right\rangle }{\left\langle
\Psi_{T}\right|  e^{-\theta H_{\text{AIM}}}\left|  \Psi_{T}\right\rangle
}.
\end{eqnarray}

We now relate the projection
formula to an artificial finite temperature ($1/\tilde{\beta}$) problem,
using  an auxiliary  Hamiltonian  $H_{T}$ whose ground state
is $\left|  \Psi_{T}\right\rangle$:
\begin{eqnarray}
\!\!\!\langle\mathcal{O}
\!\rangle_{0}\!&\!=\!&\!\lim_{\theta\rightarrow\infty}\lim_{\tilde{\beta}\rightarrow\infty}\!\frac
{\operatorname*{Tr}\left[  e^{-\tilde{\beta}H_{T}}e^{-\theta H_{\text{AIM}}/2}%
\mathcal{O}e^{-\theta H_{\text{AIM}}/2}\right]  }{\operatorname*{Tr}\left[
e^{-\tilde{\beta}H_{T}}e^{-\theta H_{\text{AIM}}} \right]  }.
\label{projection}
\end{eqnarray}
With this trick, the r.h.s.\ of 
 Eq.\ (\ref{projection}) is in a suitable form to apply (or
rederive) the Hirsch-Fye algorithm at finite values of $\tilde{\beta}$.
In the following, we will point out the necessary steps in short, 
more details will be published elsewhere \cite{Feldbacher}.
 
We discretize imaginary time into Trotter  slices of
length  $\Delta\tau$ and decouple the Coulomb interaction 
via the Hubbard-Stratonovich transformation.
Let us now restrict to single-particle trial
wave functions  $\left|  \Psi_{T}\right\rangle$
and Hamiltonians  $H_{T}$.
Then, the interaction (and the Hubbard-Stratonovich field)
is zero for the $\tilde{\beta}$ part, and we can
 combine the   propagation  of this part ($\exp[-\tilde{\beta} H_{T}]$)
with the  $H_{0}$ part of the first  $\theta$ Trotter  slice
($\exp[ -\Delta\tau H_{0}]$). 
The limit $\tilde{\beta}\rightarrow \infty$ can now
be taken analytically, leaving a problem on the interval $[0,\theta]$
discretized into $L=\theta/\Delta\tau$ steps.
As the HF-QMC, our PQMC algorithm deals with
$L\times L$ matrices for the f-electron Green function
on this interval. The updating equations for these Green functions
(if a Hubbard-Stratonovich field is changed)
are the same as that of  HF-QMC.
Choosing $H_{T}=H_{0}$, also the
starting Green function (with all Hubbard-Stratonovich fields zero)
has a similar structure: In HF-QMC it is the finite-temperature non-interacting
Green function. In 
our PQMC algorithm
it is -- because of the extra $\tilde{\beta}\rightarrow\infty$ projection --
 the zero-temperature  non-interacting
Green function $\mathcal{G}_{0}
(\tau,\tau^{\prime})$ truncated to $0\leq\tau,\tau'\leq\theta$.
But with this different object (initial Green function), 
the same program code can actually be used.

Our PQMC algorithm allows the measurement of 
 the Green function $G\left(
\tau,\tau^{\prime}\right)$ 
but also of arbitrary
correlation functions,
$\theta$-projected from 
the slater
determinant of the non-interacting AIM as a
 trial 
wave function $\left|  \Psi_{T}\right\rangle$.
For a meaningful
PQMC calculation,
 we should measure all correlation functions only on the
central ${\cal L}$ time slices. Then ${\cal P}=(L-{\cal L})/2$ time slices on the right and left
side of the measuring interval serve for projection. In the following, 
QMC results
are understood in this sense.

{\it DMFT self-consistency implementation --}
We now implement this algorithm as an impurity solver in
the DMFT self-consistency cycle. Let us consider
the half-filled Hubbard
model 
\begin{equation}
H=-t \sum_{\left\langle i,j\right\rangle\sigma }
c_{i\sigma}^{\dagger}c_{j\sigma}^{\phantom{\dagger}}+U\sum_{i}n_{i\uparrow}n_{i\downarrow},
\end{equation}
where $\langle i,j\rangle$ denotes the restriction to nearest neighbor hopping
in $d$ dimensions. We use the Bethe lattice with
 semicircular density states $N(E)=\frac{8}{\pi W^2}
 \sqrt{W^2/4-E^{2}}$. In the following, 
units are normalized to a bandwidth $W=4$.

DMFT \cite{DMFT2} maps the Hubbard model onto an
AIM with non-interacting Green function
\begin{equation}
\mathcal{G}_{0}(i\omega_{n})=[G^{-1}(i\omega_{n}) +\Sigma(i\omega)]^{-1},
\label{calG}
\end{equation}
where $\omega_n$ denote Matsubara frequencies.
The Green function $G(i\omega_{n})$ of the AIM has to be determined
self-consistently together with the self energy $\Sigma(i\omega)$.
Both are
connected by the ${\mathbf k}$-integrated Dyson
equation
\begin{equation}
G(  i\omega_{n})     =\int{\rm d}E\frac{N(E)}{i\omega-E-\Sigma
(i\omega_{n})}. \label{Dyson}
\end{equation}
Thereby, the DMFT self-consistency cycle is as follows:
\begin{equation*}
 \Sigma (i\omega_{n}) \stackrel{(\ref{Dyson})}{\rightarrow} G(  i\omega_{n}) \stackrel{(\ref{calG})}{\rightarrow}\mathcal{G}_{0}(i\omega_{n})  \stackrel{\rm PQMC}{\rightarrow} G(\tau)  \stackrel{(\ref{calG})}{\rightarrow} \Sigma (i\omega_{n}) \cdots
\end{equation*}
For the difficult task, i.e., calculating the zero temperature Green function
$G(\tau)$ of the AIM for a given 
$\mathcal{G}_{0}(\tau)$ with $\tau\in\lbrack-\theta,\theta]$, 
we use the PQMC algorithm, instead of HF-QMC \cite{Jarrell92}.
Since the PQMC Green function $G(\tau)$ is only defined for
$\tau\leq {\cal L}\Delta\tau,$ we first extrapolate 
$G\left(  \tau\right)$ to large
 $\tau$'s before Fourier transforming 
to Matsubara frequencies in the above self-consistency cycle.
To this end, we
employ the maximum entropy method, yielding the
  spectral
function $A\left(  \omega\right) $ which allows to calculate $G\left(
i\omega_{n}\right)  =\int{\rm d}\omega\frac{A\left(  \omega\right)  }{i\omega_{n}-\omega
}$ at any frequency $i\omega_{n}$.
It is important to note that,  in the end, this procedure
only serves as a fit in imaginary time. Hence, we 
do not suffer from the usual problem -- determining $A(\omega)$
from $G(\tau)$ is ill-conditioned. Only for very small $iw_n$,
corresponding to the extrapolation in imaginary time,
this procedure eventually breaks down.

\begin{figure}[ptb]
\begin{center}
\includegraphics[
width=3.0777in
]{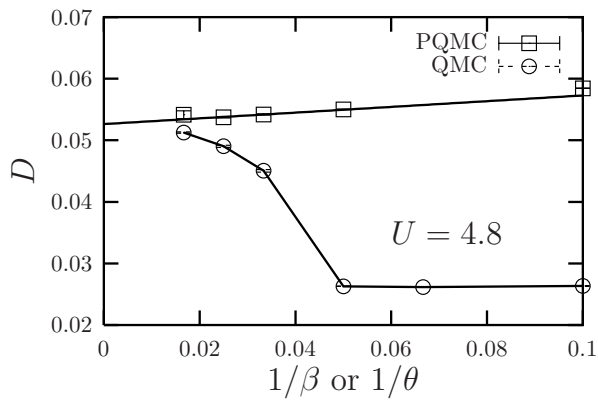}
\end{center}
\vspace{-0.3cm}
\caption{
DMFT double occupancy $D=\langle n_{\uparrow}n_{\downarrow} \rangle$
as a function of (i)  temperature $T=1/\beta $  (conventional HF-QMC; circles) 
and (ii) the  inverse projection parameter $1/\theta $ (PQMC, squares).
For  $\theta=\beta$ the numerical effort is
comparable, but the PQMC results converge much faster
and  almost linearly to the $T=0$ value. Here and in the following,
the bandwidth $W=4$ sets the energy unit.}
\label{double_Tscale_0.2}%
\end{figure}

{\it Results --}
We will now use our PQMC method to study the interaction driven Mott-Hubbard
metal-insulator transition that occurs at half filling. 
First, in Fig.\ \ref{double_Tscale_0.2} we illustrate the power of the 
PQMC method in  direct comparison with finite-$T$ HF-QMC results.
We choose equal values of
inverse temperature $\beta$ and total projection time $\theta$, having an
identical number of Trotter time slices for both calculations. Then, our
results for $\beta=\theta$ are obtained with a comparable numerical effort. 
For $U=4.8$ and at low $T$, the DMFT equations have two
solutions, where we follow the metallic solution (as soon as it exists).
In the case of DMFT calculations with finite-$T$  HF-QMC,
the  double occupancy per site $D=\langle n_{\uparrow}n_{\downarrow}\rangle$ of Fig.\
\ref{double_Tscale_0.2}  starts at high
temperatures ($\beta=10$) in the insulating phase and only after the
transition to the metallic phase around $\beta=40$ convergence to the ground
state value sets in. In contrast, the PQMC double occupancy convergences 
 much faster and extrapolates almost
linearly to the T=0 value.
For example, the $\theta=10$ PQMC result is even closer to the T=0 result
than the  HF-QMC double occupation at $\beta=30$.
Since the numerical effort grows cubically with 
 $\theta$ or $\beta$ this means that the 
PQMC is roughly  27-times faster at the same accuracy.
We attribute this behavior to the absence of
thermal fluctuations in PQMC. 

\begin{figure}[ptb]
\begin{center}
\includegraphics[
trim=0.000000in 0.000000in 0.000000in -0.039948in,
width=3.1926in
]{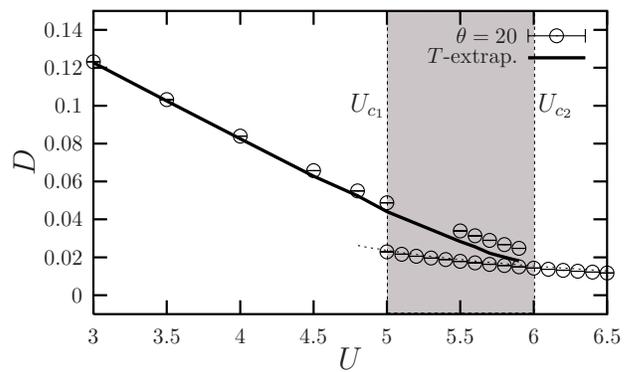}
\end{center}
\vspace{-0.5cm}
\caption{Double occupancy $D=\langle n_{\uparrow}n_{\downarrow}\rangle$ as a function of
interaction $U$ for both metallic and insulating DMFT solutions. In the
interval $5.0\approx U_{c_{1}}<U<U_{c_{2}}\approx 6.0$ both solutions coexist (shaded
region). Circles: $\theta =20$;
thick line: extrapolation to  $\theta \rightarrow \infty$;
dashed line: strong coupling expansion of \cite{Bluemer04}.}%
\label{double_b20_0.2}%
\end{figure}

In the following, we present our results for the Mott-Hubbard metal-insulator
transition, starting with the double occupation
in Fig.\ \ref{double_b20_0.2}.  All calculations are done at a fixed $\Delta\tau=0.2$. Upon
increasing $U$, we systematically follow the metallic solution until at
$U_{c_{2}}\approx 6.0$ the metallic quasiparticle peak disappears. For the
insulating solution on the other hand, 
we decrease the interaction until the insulator becomes
unstable below $U_{c_{1}} \approx 5.0$. Within the
uncertainty of roughly 0.1-0.2,
our coexistence region  $U_{c_{1}%
}<U<U_{c_{2}}$ compares well with NRG results ($U_{c_{1}}=4.78$,
$U_{c_{2}}=5.88$ \cite{NRGT=0,NRGT}). 
Already the $\beta=20$ results (circles) for the
double occupancy yield a good estimate for the low temperature coexistence
region. In order to resolve the increasingly sharp quasiparticle peak in the
vicinity of the critical point we also used longer projection times $\theta$
and the thick line represents a linear extrapolation for three values $\theta
=20,30,40$. Also plotted is the result from a 10th order
perturbation expansion \cite{Bluemer04} in  $t/U$,
which agrees well with our results for the insulator.
Our results are consistent with a linear dependence 
$(D_{\text{met}}-D_{\text{ins}})\propto(U-U_{c_2})$ which may be integrated to a groundstate
energy $(E_{\text{met}}-E_{\text{ins}})\propto(U-U_{c_2})^2$ corresponding to a second order 
transition at $U_{c_2}$ \cite{Moeller95}.

\begin{figure}[ptb]
\begin{center}
\includegraphics[
height=2.6472in,
width=2.8902in
]{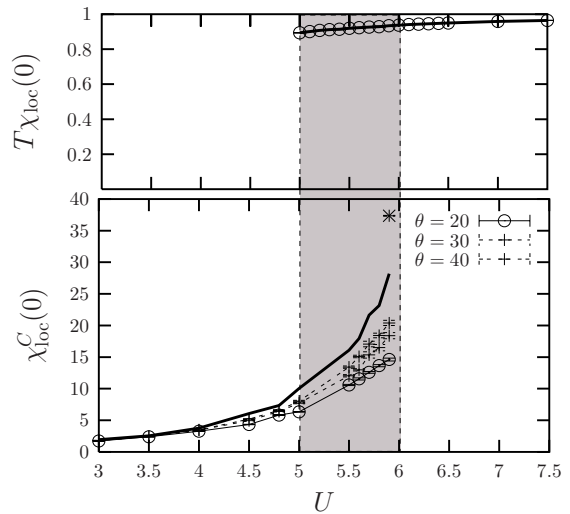}
\end{center}
\vspace{-0.3cm}
\caption{Static local spin susceptibility $T\chi_{\text{loc}}(0)$ vs.\ $U$
for the insulating (upper panel) and metallic DMFT solution
(lower panel). For the metal, $\chi_{\text{loc}}(0)$
diverges as we approach $U_{c_{2}}$ from below. 
Since we plot the truncated integral $\chi_{\text{loc}}^{C}$,
the divergence is cut off at 
$\chi_{\text{loc}}^{C}=37.4$ (star).}%
\label{chi}%
\end{figure}

Fig.\ \ref{chi} plots the local spin susceptibility $
%
T\chi_{\text{loc}}\left(  0\right)  =\lim\limits_{\theta\rightarrow\infty
}\frac{1}{\theta}\int_{0}^{\theta}\left\langle S^{z}\left(  \tau\right)
S^{z}\right\rangle d\tau$, and
actually represents the first DMFT analysis of this
quantity through the zero temperature Mott-Hubbard metal-insulator transition
(at finite-$T$ it was studied in \cite{Rozenberg94}).
 In the top panel, we observe the local moment
behavior of the insulating solution without any sign of criticality at
$U_{c_{2}}$. On the other hand approaching $U_{c_{2}}$ from the metallic side,
$\chi_{\text{loc}}$ is expected to diverge, reflecting the critical behavior
of a Fermi liquid with a vanishing quasiparticle weight $Z$ \cite{Moeller95}.
Therefore the second order character of the phase transition at $U_{c_2}$ can only be observed 
for $U<U_{c_2}$.
For the metallic phase,
we integrate the  spin correlation function 
up to a maximal cutoff  $C$:
$\chi_{\text{loc}}^{C}\left(  0\right)  =\int_{0}^{C}\left\langle S^{z}\left(
\tau\right)  S^{z}\right\rangle d\tau$.
 The thick
line is an extrapolation of
 $\chi_{\text{loc}}^{C}$ for a fixed cutoff $C=40$ to 
$\theta\rightarrow \infty$, i.e., to the
groundstate value. At the critical point, we see indications for a divergence
of $\chi_{\text{loc}}$
which for the cutoff quantity $\chi_{\text{loc}}^{C}$ means approaching
the star ($\chi_{\text{loc}%
}^{C}=37.4$, the value of the insulating solution).

\begin{figure}[ptb]
\begin{center}
\includegraphics[
height=1.9936in,
width=3.2623in
]{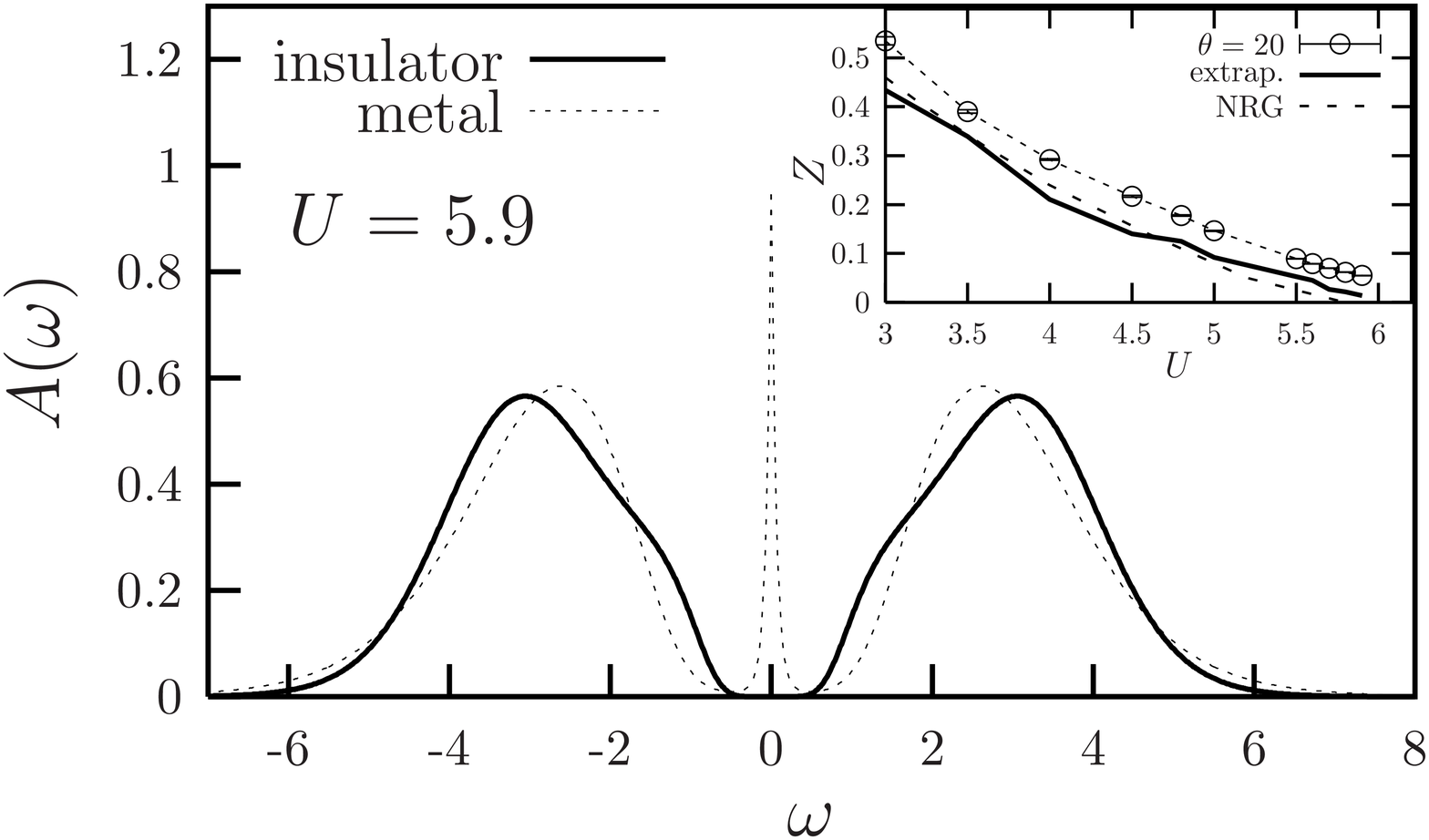}
\end{center}
\caption{Spectral functions $A\left(  \omega\right)  $ for the insulating
(line) and metallic solution (dashed) close to the phase transition at
$U=5.9$). Inset: Quasiparticle weight $Z$ vs.\ $U$ 
for 
$\theta=20$ (circles) and extrapolated to $\theta\rightarrow \infty$
(thick line), compared with NRG  \cite{NRGT=0} (dashed line).
 }%
\label{aom_u5.9_m40_i20+z}%
\end{figure}

Finally, we present in Fig.\ \ref{aom_u5.9_m40_i20+z}
the spectra  at $U=5.9$, close to the phase
transition. The insulating solution ($\theta=20$) has a
pronounced charge gap of around $1.2$ between two Hubbard bands. 
The metallic solution
shows an additional quasiparticle peak which
is pinned to the non-interacting value,
 a consequence of Fermi liquid theory at $T=0$ \cite{MuellerHartmann89c}.
The 
quasiparticle peak for a fixed $\theta=40$ has width of $0.05$
(Note, that the low energy
accuracy is limited by $1/\theta$ and the statistical quality of QMC data.).
We determine the quasiparticle weight  by 
$Z=\left(  1-\frac{\operatorname{Im}%
\Sigma\left(  i\omega_{1}\right)  }{\omega_{1}}\right)^{-1}$ from
the first Matsubara frequency. In principle, $\omega_{1}$ can
be chosen to be arbitrarily small, but controlled values
for  $\Sigma\left(i\omega_{n}\right)$ are restricted by the
projection length $1/\theta$ and the maximum entropy extrapolation fit.
 The inset in
Fig.\ \ref{aom_u5.9_m40_i20+z} shows that at $\theta=20$ the weight $Z$ is
still too large. But  after extrapolating $Z$ from $\theta=20,30,40$
to $\theta\rightarrow\infty$, we find
agreement with the NRG results \cite{NRGT=0}.

{\it  Summary --} We 
have introduced a novel projective (PQMC) algorithm for impurity
models, which is on the technical front closely related to the standard
Hirsch-Fye algorithm. The numerical effort for both algorithms is identical,
but the convergence to the ground state is dramatically improved by PQMC. 
Using PQMC as an impurity solver for
DMFT, we were able to obtain accurate zero temperature
results for the Mott-Hubbard metal-insulator transition.
There has been a strong controversy about the nature
of this Mott-Hubbard transition within DMFT,
in particular, whether  two solutions (metallic and insulating)
coexist in an interval $U_{c1}\!<\!U\!<\!U_{c2}$ or not.
While the coexistence is by now generally accepted,
NRG calculations \cite{NRGT=0} represent the only unassailable 
zero-temperature confirmation of the coexistence scenario \cite{DMFT2}.
Since other numerical  results at $T=0$  \cite{RDA} and 
analytical arguments \cite{AnalyticArg} contradict this scenario,
an independent numerical study is certainly in order.
Using PQMC, we  have provided for such  a mandatory reconfirmation, 
and beyond it we
presented results for the double occupancy and  local susceptibility.

{\it Outlook --} 
In recent years,  there have been two main lines of DMFT developments: 
towards realistic calculations such as LDA+DMFT \cite{LDADMFT}
and towards the inclusion of more and more (non-local) correlations
by means of
cluster DMFT calculations \cite{ClusterDMFT}.
These approaches require additional orbitals and DMFT sites, respectively,
making NRG prohibitively CPU-expensive 
since the effort scales exponentially with more orbitals and sites.
In contrast, 
our PQMC algorithm only scales quadratically or cubically,
and hence opens the door
to accessing low temperatures within realistic approaches
or model-oriented calculations with non-local correlations.

We thank N.\ Bl\"umer for discussions.
This work was supported in part by the Deutsche Forschungsgemeinschaft
through the Emmy Noether program.

%
%
%
%
%
%
%
%
%
%
%
%
%
%
%
%
\end{document}